\documentclass[aps,prl,reprint,twocolumn,superscriptaddress]{revtex4}

\usepackage{units} % ADDED LJK
\usepackage{graphicx}
\usepackage{amssymb,amsmath}
\usepackage{graphics}
\usepackage{moreverb}
\usepackage{epsfig}
\usepackage{subfigure}
\usepackage{hangcaption}
\usepackage{txfonts}
\usepackage{palatino}
\usepackage{ifthen}
\usepackage[usenames]{color}
\usepackage{placeins}
\usepackage{upgreek}
\usepackage{algorithmic}
\usepackage{hyperref}
\begin{document}                  % DO NOT DELETE THIS LINE

\title{X-ray analog pixel array detector for single synchrotron bunch time-resolved imaging}

\author{Lucas J. Koerner}
\email{ljk29@cornell.edu}
\affiliation{Dept. of Physics, LASSP, Cornell University, Ithaca NY, USA}
%\author[a]{Mark W.}{Tate}
\author{Sol M. Gruner}
\email{smg26@cornell.edu}
\affiliation{Dept. of Physics, LASSP, Cornell University, Ithaca NY, USA}
\affiliation{Cornell High Energy Synchrotron Source, CHESS}
\begin{abstract}
Dynamic x-ray studies may reach temporal resolutions limited by only the x-ray pulse duration if the detector is fast enough to segregate synchrotron pulses.  An analog integrating pixel array detector with in-pixel storage and temporal resolution of around \unit[150]{ns}, sufficient to isolate pulses, is presented.  Analog integration minimizes count-rate limitations and in-pixel storage captures successive pulses.  Fundamental tests of noise and linearity as well as high-speed laser measurements are shown.  The detector resolved individual bunch trains at the Cornell High Energy Synchrotron Source (CHESS) at levels of up to $3.7\times 10^3$ x-rays/pixel/train.  When applied to turn-by-turn x-ray beam characterization single-shot intensity measurements were made with a repeatability of 0.4$\%$ and  horizontal oscillations of the positron cloud were detected.  This device is appropriate for time-resolved Bragg spot single crystal experiments.
\end{abstract}
\maketitle

\section{Introduction}
Synchrotrons are pulsed x-ray sources that may be exploited for time-resolved experiments.  Isolation of the synchrotron pulses allows for dynamic studies limited by the x-ray pulse width of around \unit[50]{ps}~\cite{Decamp:2005}.  Mechanical chopper systems have been designed to transmit one x-ray pulse at a duty cycle of around \unit[1]{kHz}~\cite{cammarata:2009}.  When synchronized to a laser pump, choppers allow conventional x-ray detectors to be used for ultrafast experiments.  However, detectors that isolate successive x-ray pulses from the synchrotron under control of electronic gate signals ease the experimental design and allow for a wider range of experiments.\\
\indent
Single bunch pump-probe experiments have used point avalanche photodiodes (APD) coupled to counting external electronics to measure up to one x-ray per pulse \cite{Baron:1997}.  The output of an APD may also be processed by analog electronics and then digitized to isolate successive x-ray pulses with a signal capacity over 500 x-rays per pulse~\cite{Cheong:2004}.  Pilatus, a photon counting pixel array detector (PAD), was applied to single bunch experiments at the Advanced Photon Source (APS)~\cite{Ejdrup:2009}.  To do so, the in-pixel counter of Pilatus was gated by an external control signal for isolation of single pulses at a count-rate of one x-ray per pixel per pulse.  Image intensifiers with a gate mode as short as \unit[3]{ns} have been included in the optical chain of an x-ray CCD detector to isolate a single x-ray pulse~\cite{nuske:2010}.  The detectors described above have limitations for high-speed experiments that analog integrating pixel array detectors may address.  APDs are single pixel, digital PADs are limited in count-rate and subsequently accuracy per shot, and the two-dimensional area detectors discussed above cannot isolate and record successive synchrotron pulses.  \\
\indent
Analog integrating pixel array detectors that isolate synchrotron pulses will allow for new paradigms of single bunch x-ray experiments and more efficiently acquire data from conventional pump-probe configurations.  An analog integrating PAD is shown in this paper to measure 1000s of x-rays per pixel per pulse for single-shot Poisson limited accuracies at each pixel of $3\%$ or better.  In-pixel frame storage captures a number of images in rapid succession before detector read out to particularly benefit the study of samples with spontaneous non-reversible changes~\cite{Trenkle:2008}.\\
\indent
% now very specific about the device
The detector consists of a 16$\times$16 pixel CMOS integrated circuit hybridized to a high resistivity silicon detector.   The hybrid was combined with support electronics and flexible Field-Programmable Gate Array (FPGA) control and acquisition to create an x-ray camera.  The support electronics and FPGA code allowed for a minimum exposure time of \unit[30]{ns} with \unit[10]{ns} resolution, a \unit[600]{$\mu$s} readout, and buffering for 8,100 frames before a transfer to hard-disk was required.  The fundamental detector metrics of noise and linearity are presented.  Pulsed laser and x-ray synchrotron experiments show the PAD to have sufficient speed to isolate synchrotron pulses at a spacing of around \unit[150]{ns}.  This time resolution is appropriate for the \unit[176]{ns} bunch separation in 16 bunch mode at the European Synchrotron Radiation Facility (ESRF) and the \unit[153]{ns} bunch separation in a standard operating mode at the APS.
\section{Device description}
\begin{figure}[]
\includegraphics[width=3.5in]{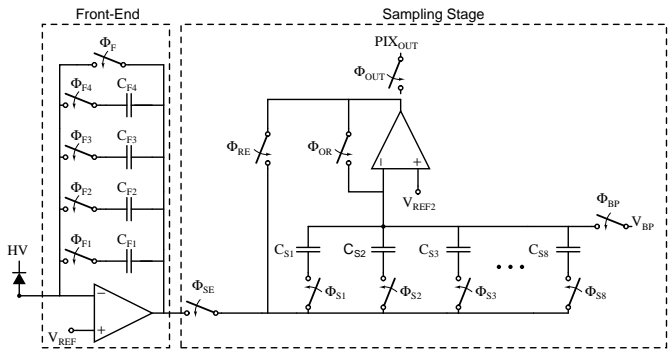}
\caption{Simplified pixel schematic that differentiates the front-end stage and the sampling stage.  The reversed biased diode represents the high resistivity detector layer.}
\label{fig:schematic}
\end{figure}

\begin{figure*}[]
\includegraphics[width=5.5in]{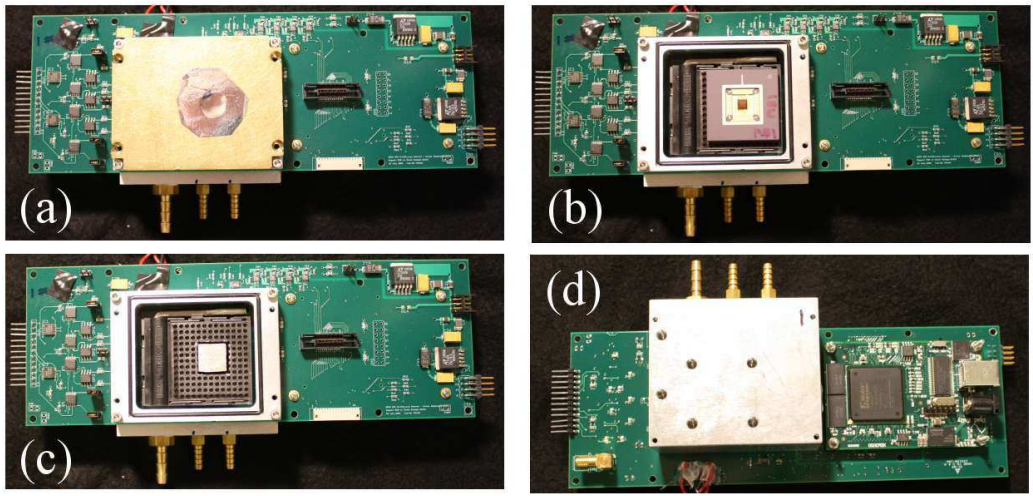}
\caption{Photographs of the support PCB and enclosure from different angles and at different
levels of construction.  (a) The system entirely assembled from the
top-side.  An x-ray transparent aluminized mylar window covered a
hole in the top brass cover.  The three barbed
hose-fittings are a vacuum port and connections
for cooling water. (b) The enclosure with the brass
cover removed to reveal the PGA and chip mounted in the ZIF socket.
(c) The chip is removed to show the copper cold-finger that protrudes through the ZIF socket. (d) PCB photographed from the back-side so that the control FPGA board is shown on the right.  The support PCB measured 8.0" $\times$ 2.85".}
\label{fig:enclosureAndPCB}
\end{figure*}

The CMOS readout application specific integrated circuit (ASIC) is a 16$\times$16 pixel array with $\unit[150]{\mu m}$ pitch designed in TSMC \unit[0.25]{$\mu$m} process and submitted to MOSIS as part of a multi-project wafer run.  The circuit layout allowed for a smaller pixel pitch but \unit[150]{$\mu$m} was chosen to allow for hybridization to available detector layers.  All NMOS transistors were designed using either an enclosed layout~\cite{Anelli:1999} or radiation hardened linear techniques~\cite{Snoeys:2002}.  The array was divided in half; one side matched an earlier prototype while the other side implemented a few slight design modifications to address issues revealed during testing of an earlier prototype chip (particularly radiation robustness)~\cite{Koerner:2009}.  The chip has four analog output ports. \\
\indent
The pixel electronics, shown as a simplified schematic in Fig.~\ref{fig:schematic}, are similar to the analog integrating approach with in-pixel storage used in past detectors~\cite{Rossi:1999,Ercan:2006}.  The design has been updated with an emphasis on speed for bunch segregation at the \unit[150]{ns} level.  The in-pixel amplifiers use a differential architecture for radiation hardness and to limit systematic effects at high input x-ray flux.  The front-end capacitors ($C_{F1}$ - $C_{F4}$ in Fig.~\ref{fig:schematic}) may be re-addressed and signal added without reading out the device.  This addition of signal may occur after either acquisition into a different capacitor or electronic shuttering of the x-ray signal.  Each in-pixel frame may thus be built from temporally separated acquisition windows which allows for in-pixel averaging.  A distinct temporal window is referred to as an accumulation. \\
\indent
Switches $\Phi_{F1}-\Phi_{F4}$ may be held fixed to bypass accumulation.  In this case, referred to as flash-mode, the pixel captures eight sequential images that are stored on $C_{S1} - C_{S8}$ before readout.  The front-end conversion gain, when accumulation is not used, is set by the configuration of switches $\Phi_{F1}-\Phi_{F4}$ and adjustable by a factor of up to 6.5.  Pixels have four accumulation elements ($C_{F1}$ - $C_{F4}$) and eight storage elements ($C_{S1}$ - $C_{S8}$).
\indent
\\
\indent
ASICs were hybridized to high-resistivity silicon detector layers.  The detector layers were n-type \unit[500]{$\mu$m} thick layers fabricated by SINTEF (Oslo, Norway) with a gold pad metallization.  Pixel p+ implants were at the bump-bonded side and an aluminized n+ ohmic contact was on the x-ray incident side for the application of a bias to deplete the thickness of the sensor.  A substrate of $\unit[7.5]{k\Omega \cdot cm}$ resistivity and a reverse bias of \unit[250]{V} gives an anticipated hole collection time of \unit[25]{ns}.  With a larger overbias the minimum collection time, constrained by the hole saturated velocity, is \unit[8.3]{ns}. \\
\indent
To bump-bond the detector layers to ASICs a process compliant with diced chips returned from MOSIS multi-project wafer runs was required.  Gold stud bump bonding with conductive adhesives was used since it is compliant with small diced chips (Polymer Assembly Technologies, Research Triangle Park, NC).  Gold stud bumps were attached to the aluminum pad on the ASIC chip by intentionally removing the wire from a thermosonic wirebond to leave
only the ball-bond.  A conductive polymer epoxy bump was placed on the pads of the detector through a stencil screen~\cite{clayton:2003TNS}.\\
%\subsection{FPGA readout}
\indent
An FPGA board (XEM3050 Opal Kelly,
Portland, OR) with a Xilinx Spartan-3 FPGA was
used to send digital control signals to the PAD, to buffer data from analog-to-digital converters, and to transfer data to a computer.  The FPGA board featured a
USB 2.0 interface for download of the FPGA configuration and for writing of exposure commands from a computer to registers in the FPGA.  The FPGA system continuously read up to 8,100 frames from the PAD by buffering onto a memory chip before a USB transfer to a computer was required.  The exposure and reset times were determined by the FPGA and programmable from \unit[30]{ns}-\unit[21.4]{s} with a master FPGA clock of \unit[10]{ns}.\\
%\subsection{Support board and clam-shell cryostat}
\indent
Detectors were packaged in ceramic pin grid array (PGA) carriers.  PGAs were mounted into a zero-insertion-force (ZIF) socket with a central hole that allowed a copper heat-sink to contact the back-side of the PGA.  The temperature of the copper heat-sink was regulated by a thermoelectric device and an RTD sensor.   A support printed circuit board (PCB), detector enclosure, and the FPGA board are shown in Fig.~\ref{fig:enclosureAndPCB}.  The enclosure allowed for evacuation of the detector environment to prevent condensation when the detector was cooled.  Electrical signals were transmitted into the enclosure through inner-layers of the PCB.
\section{Fundamental metrics, laser measurements, and radiation hardness}
\subsection{Fundamental detector specifications}
The RMS equivalent noise charge (ENC) of the modified half of a hybridized PAD for flash-mode operation was measured at $\unit[18]{^\circ C}$ to be 1000, 1445, 2800, 3415, and \unit[5500]{e$^{-}$} for capacitive feedback configurations of 300, 466, 966, 1200, and \unit[1966]{fF}, respectively by engaging combinations of switches $\Phi_{F1}$ to $\Phi_{F4}$.  Therefore, a signal-to-noise of 2.2 for the detection of a single 8~keV x-ray is possible at the highest front-end gain setting.  The noise of the other half, due to differences in the sampling stage, was around $20\%$ greater.  The PAD was measured to swing \unit[1.88]{V} with a non-linearity of less than $1\%$.  At the lowest gain the per-pixel well-depth for less than $1\%$ non-linearity is 10,480 x-rays of 8~keV energy.
\subsection{High dose-rate laser measurements}
\begin{figure}[] \centering
\includegraphics[width=3.5in]{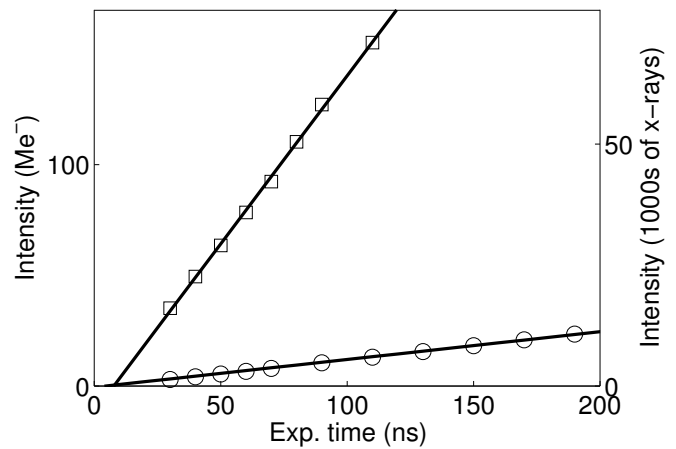}
\caption{Detector linearity at nanosecond exposure times tested with a laser source.  $\square$ used the laser unattenuated. $\circ$ used a neutral-density filter of optical density 1.0.  Solid lines are a linear fit to the measured points.}
\label{fig:linearity_laser}
\end{figure}

\begin{figure}[] \centering
\includegraphics[width=3.5in]{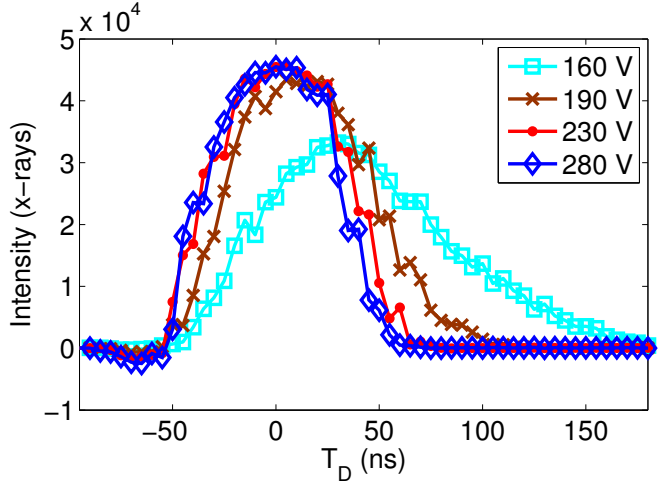}
\caption{Integrated intensity in a detected laser spot (in equivalent \unit[8]{keV x-rays}) for a laser pulse of \unit[40]{ns} duration versus the delay between PAD acquisition and the pulse.  The experiment was repeated at multiple values of the detector layer bias (HV in Fig.~\ref{fig:schematic}), indicated in the legend.  Lines guide the eye.}
\label{fig:laserspeed}
\end{figure}

The PAD linearity was studied at high input photocurrents using a laser of \unit[633]{nm} wavelength and \unit[0.5]{mW} maximum output power (Melles Griot, Albuquerque, NM, Model 25-LHP-213-249).  For efficient visible detection the aluminum on the x-ray entrance side was removed from one of the detector chips.  Fig.~\ref{fig:linearity_laser} shows the integrated intensity in the laser spot versus the exposure time acquired at two levels of incident intensity.  This result confirms \unit[10]{ns} resolution of the FPGA timing circuits.\\
\indent
The total photocurrents measured were \unit[243]{$\mu$A} and \unit[20]{$\mu$A} for direct laser illumination and attenuation with a filter of optical density 1.0.  The maximum per pixel photocurrent was \unit[12]{$\mu$A} without attenuation, which is comparable to the static bias current of the front-end amplifier, and the response remains linear.  The photocurrent produced by the unattenuated laser is equivalent to the signal from an \unit[8]{keV} x-ray flux of \unit[$7\times10^{11}$]{x-rays/s}, around a factor of 1000 greater than what is accessible with laboratory x-ray sources.  The largest single pixel flux, equivalent to \unit[$3.4\times10^{10}$]{x-rays} of 8 keV energy, exceeded the maximum pixel count-rate  (\unit[10]{MHz}) of area digital PADs by over 3,000.  The intercepts of Fig.~\ref{fig:linearity_laser} were near to zero; the deviation may be because the effective exposure time was slightly different than what was programmed into the FPGA.
\\
\indent
A laser (Newport Corp., Irvine, CA, Model LQA635-03C) with analog modulation at up to \unit[20]{MHz}, a wavelength of \unit[635]{nm}, and a maximum output power of \unit[3]{mW} was used to illuminate the detector with a \unit[40]{ns} pulsed input to test high-speed collection.  A delay generator (Stanford Research Systems, Sunnyvale, CA, Model DSG 535) controlled the delay between laser pulse and PAD acquisition.  Data was acquired with a \unit[90]{ns} exposure time, $C_F$ = \unit[1000]{fF}, and with a front-end amplifier dissipation of \unit[12.3]{$\mu$A}.  Fig.~\ref{fig:laserspeed} shows the integrated intensity in the detected laser spot versus the delay between the laser pulse and the start of the PAD exposure window ($T_D$) acquired at multiple values of the detector layer bias.  At larger $T_D$ the PAD was started later with respect to the laser.  At bias voltages of 280 and \unit[230]{V} the integrated intensity is flat versus $T_D$ for $T_D$ greater than \unit[-10]{ns}.  This indicates that the signal was completely measured in the \unit[90]{ns} exposure window when the PAD was started \unit[10]{ns} before the laser was pulsed.  The signal was also completely measured when the PAD start was coincident with the laser pulse.
\subsection{Accumulation studies}
\begin{figure}[]
\includegraphics[width=3.5in]{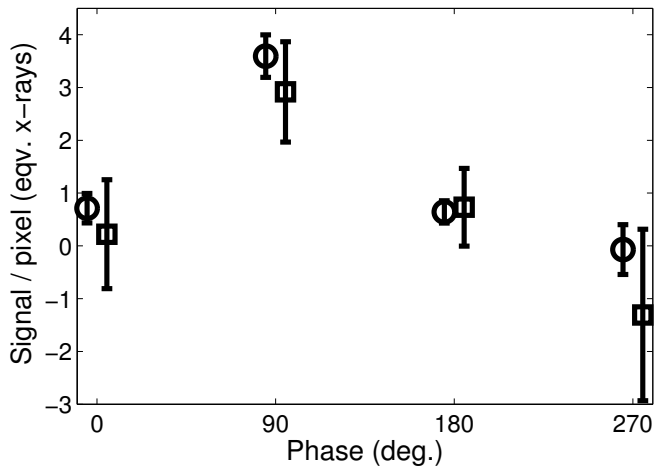}
\caption{Signal per pixel (in equivalent \unit[8]{keV} x-rays) captured at four phases of an oscillatory stimulus.  $\square$ shows data averaged in post-processing.  $\circ$ shows data acquired using the accumulation functionality, i.e. averaged in pixel.  The $\circ$ ($\square$) are offset horizontally by $-5^{\circ}$ ($+5^{\circ}$) for easier visualization. }
\label{fig:accum_demo}
% 20091023_NB12pg38
\end{figure}

The noise per accumulation was studied for multiple values of feedback capacitance.  The noise measured was differentiated into two sources by a fit to $\sigma(N)^2 = \sigma_{F}^2 + N\sigma_{acc}^2$, where $\sigma$ is the total noise measured, $\sigma_F$ is the fixed noise, $\sigma_{acc}$ is the noise added per accumulation, and $N$ is the number of accumulations.  The noise per accumulation was found to follow the form $(\sqrt{kTC_{F}}+\sqrt{kTC_{IN}})/C_{F}$ where $C_{F}$ is the feedback capacitance and $C_{IN}$ is the input capacitance due to the detector layer and input amplifier.  Table~\ref{table:accum_noise} shows the fixed read-noise and noise per accumulation extracted from the data fits from the modified half of a hybridized detector.  Also shown is the number of accumulations possible before the added accumulation noise matches the fixed read-noise.\\
\indent
\begin{table}
\caption{Noise measured from accumulations differentiated as fixed noise and noise per accumulation.  The fourth column shows the number of accumulations until accumulation noise matches the fixed noise.}\label{table:accum_noise}
\begin{tabular}{llll}
  \hline
  $C_F$ (fF) & $\sigma_F$ ($\mu$V) &$\sigma_{acc}$ ($\mu$V) & $N$ for $\sigma_{F}^2$ = N$\sigma_{acc}^2$ \\
  \hline
  300 & 550 & 186 & 9 \\
  466 & 496 & 133 & 14 \\
  500& 490 & 127 & 15 \\
  700 & 473 & 96 & 24 \\
  1666 & 452 & 41.7 & 117 \\
  \hline
\end{tabular}
\end{table}
\indent
The accumulation functionality allows measurement of a repetitive signal with less noise than other methods by allowing for in-pixel averaging (see the fourth column of Table~\ref{table:accum_noise}).  To demonstrate this the detector was illuminated with a laser (Coherent Laser, Santa Clara, CA, Lab Laser MVP/VLM2) modulated with a sinusoid at \unit[25]{kHz}.  The light intensity was attenuated by 10,000 using a neutral density filter for low signal level imaging.  Four accumulation elements captured the intensity at four phases of the oscillation: $0^{\circ}$, $90^{\circ}$, $180^{\circ}$, $270^{\circ}$ with  the laser at full intensity at $90^{\circ}$ and off at $270^{\circ}$.  Two methods of imaging this scene were compared.  One technique captured each phase of the oscillation once with an exposure time of \unit[200]{ns} and then the image was read out.  Fifteen of these images were averaged to form a composite image.  The second technique used the four accumulation elements to capture each phase of the oscillation with a \unit[200]{ns} exposure window fifteen times before the detector image was read out.  For this second technique, the intensity was calculated from one single image rather than an average.  In both cases the total exposure time for the capture of each phase of the oscillation was $15\times\unit[200]{ns} = \unit[3]{\mu s}$.  The first technique used post-processing averaging while the second technique averaged in-pixel.  The integrated intensity measured, in units of \unit[8]{keV} x-rays per pixel, along with error-bars of $\pm$one-sigma are shown in Fig.~\ref{fig:accum_demo}.  In this experiment the number of accumulations was such that the added accumulation noise is near to that of a single readout.  Hence, the noise increase expected from fifteen image reads compared to a single read and fifteen accumulations is: $\sqrt{15/2} = 2.7$.  The ratios of the noises measured were 2.7, 2.4, 2.5, and 3.5, in reasonable agreement with expectations.
\subsection{Radiation hardness}
Radiation robustness was evaluated by dosing an ASIC without a bump-bonded detector layer held at \unit[-24]{C} at a rate of \unit[2]{Gy(Si)/sec}.  X-rays were produced by a rotating anode source (Enraf Nonius, Model FR571, Bohemia, NY) operated at \unit[40]{kV} and \unit[50]{mA} with multilayers (Osmic model CMF15-165Cu8, Troy, MI) to select \unit[8]{keV} radiation.  The chip was irradiated up to \unit[600]{kGy(Si)} over the course of three days.  With a \unit[500]{$\mu$m} thick silicon detector layer for protection a dose of \unit[600]{kGy(Si)} at the readout ASIC is reached after exposure to a flux of \unit[3$\times 10^{11}$]{x-rays/s/mm$^2$} for \unit[94]{hours} at \unit[8]{keV} ($\unit[\sim \! 800]{MGy(Si)}$ at the detector diode layer) or \unit[77]{minutes} at \unit[12]{keV} ($\unit[\sim \! 5.5]{MGy(Si)}$ at the detector diode layer).  The half of the chip that was modified for improved radiation hardness remained functional up to \unit[600]{kGy(Si)}.  The most significant effect induced by accumulated dose was a manageable increase in the subthreshold leakage of the transistor switches that isolate the storage element capacitors.
\section{Synchrotron measurements}
\subsection{High flux studies}
\begin{figure}[] \centering
\includegraphics[width=3.5in]{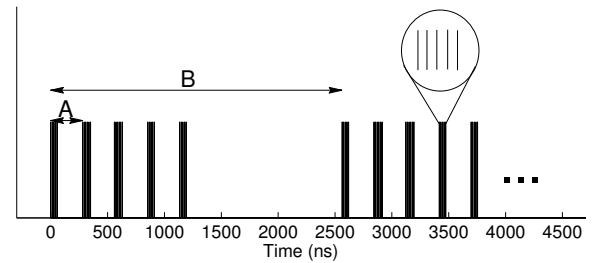}
\caption{The five trains of bunches as each circulates the synchrotron at CHESS twice.  Each individual vertical bar is a bunch of positrons.  The trains, from left to right, are referred to as 1-5.  Trains 2 and 3 had six bunches and trains 1, 4, and 5 had five bunches.  The front-to-front train spacing was \unit[280]{ns} (labeled by A).  The revolution period was \unit[2563.2]{ns} (labeled by B).  The time between bunches in a train was \unit[14]{ns}.}
\label{fig:fillpattern}
\end{figure}

\begin{figure}[] \centering
\includegraphics[width=3.5in]{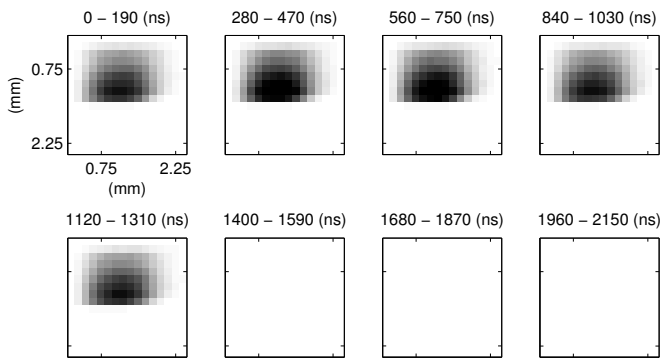}
\caption{Eight images, one from each in-pixel storage element, acquired in a single read out to illustrate the synchrotron fill-pattern. The linear gray-scale is set from -1 x-ray (white) to $3.7\times10^3$ x-rays
(black).  The time-window, \unit[190]{ns} in duration, covered
by each in-pixel storage element is labeled above the image.
Time-window 0-\unit[190]{ns} captured train 1, trains 2-5 were captured in the next
successive time-windows, and the last three time-windows captured
the period without x-rays (1200 - \unit[2560]{ns} in
Fig.~\ref{fig:fillpattern}).  The sharp edge at the bottom of the spot is hypothesized to be due to vignetting from upstream optical elements.}
\label{fig:fillpatternPAD}
\end{figure}

\begin{figure}[] \centering
\includegraphics[width=3.5in]{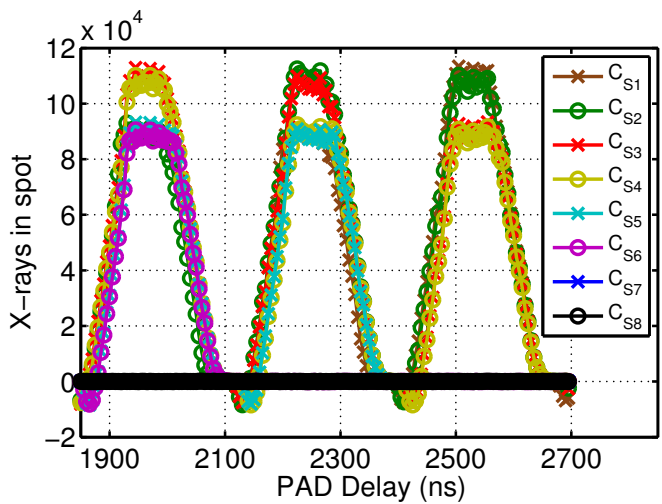}
\caption{Integrated signal versus the PAD delay with respect to the synchrotron timing with a train imaged by successive in-pixel storage elements.  The exposure time was \unit[190]{ns} and the time between frames was \unit[90]{ns}.  $C_{S7}$ and $C_{S8}$ measured around zero at all PAD delays and are indistinguishable in the figure.  Lines guide the eye.}
\label{fig:everyTrainvsDelay}
\end{figure}

The high flux x-ray performance of the detector was tested at G3 hutch of the Cornell High Energy Synchrotron Source (CHESS).  CHESS G-line receives x-ray radiation from positrons via a 49 pole wiggler.  X-rays of \unit[8.6]{keV} energy were selected by a W/B$_4$C multilayer monochromator with energy bandpass $\Delta E/E=2.1\%$. G3 receives a flux of up to \unit[$5\times10^{13}$]{x-rays/sec/mm$^2$}. In the hutch a pair of slits was used to reduce the size of the beam to around \unit[1]{mm$^2$}.  An aluminum disc with different thicknesses at each position of rotation was used to attenuate the x-ray beam when necessary.  A fast shutter with $\simeq\!\unit[5]{ms}$ opening and closing times was used to limit x-ray exposure to the detector (Uniblitz/Vincent Associates, Rochester, NY).  The synchrotron timing signal triggered a delay generator which subsequently triggered the PAD.  The time from arrival of the synchrotron trigger to release of the trigger to the PAD (referred to as $T_D$ or PAD delay) and the repetition rate of exposures were adjusted with the delay generator.  The detector layer was biased to \unit[290]{V} and the PAD temperature was stabilized at \unit[-15]{$^{\circ}$C} or \unit[-24]{$^{\circ}$C}.  PAD readout required around \unit[600]{$\mu$s}. \\
\indent
The synchrotron positron fill-pattern during this experiment is shown in the schematic of Fig.~\ref{fig:fillpattern} and demonstrated with a detector readout in Fig.~\ref{fig:fillpatternPAD}.  Five bunch trains circulated the ring with a front-to-front spacing of \unit[280]{ns} (A in Fig.~\ref{fig:fillpattern}).  The trains contained five or six bunches separated by \unit[14]{ns}.  The intra-train bunch spacing was too short to resolve so the trains were used as a stand in for a bunch. \\
\indent
Fig.~\ref{fig:fillpatternPAD} shows the capture of the 2nd and 3rd
trains contained more signal than that of the 1st, 4th, and 5th
trains.  The fill-pattern had six bunches in the 2nd and 3rd trains and five bunches in the other trains.  For the acquisition of Fig.~\ref{fig:fillpatternPAD}, the detector measured an
integrated intensity of \unit[$1.75\times10^5$]{x-rays} for the two bright trains and \unit[$1.44\times10^5$]{x-rays} for the other trains (a ratio which approximately matched that of the number of bunches: $1.22 \cong 6/5$).
\indent
The experiment displayed in Fig.~\ref{fig:fillpatternPAD} was repeated while the PAD delay ($T_D$) was adjusted with respect to the synchrotron timing to study the dependence of the PAD response upon the time within the exposure window that the x-rays arrived.  As the PAD delay was increased the x-rays arrived earlier in the exposure window.  Fig.~\ref{fig:everyTrainvsDelay} shows the integrated intensity of each storage element from a single pixel plotted versus the PAD delay.  The group of peaks at smallest PAD delay show that $C_{S2}$, $C_{S5}$, and $C_{S6}$ measured the lower intensity bunch trains (1, 4, and 5); $C_{S3}$ and $C_{S4}$ measured the high intensity trains (2 and 3); and $C_{S1}$, $C_{S7}$, and $C_{S8}$ were not illuminated. The duration of the rising and falling edges and the flat-tops of the pixel response in Fig.~\ref{fig:everyTrainvsDelay} provide information about the speed of the detector response.  For trains with six bunches the rising, flat-top, and falling edge durations were \unit[100]{ns}, \unit[60]{ns}, and \unit[100]{ns}, respectively. For the trains with five bunches the rising, flat-top, and falling edge durations were \unit[90]{ns}, \unit[75]{ns}, and \unit[85]{ns}, respectively.  The total time that signal was detected was \unit[250]{ns} for the five-bunch trains and \unit[260]{ns} for the six-bunch trains, which is approximately the sum of the exposure time, x-ray train duration, and the detector collection time which indicates that the readout ASIC electronics did not considerably slow the response.  Trains are clearly separated and trains of five bunches and of six bunches are distinguishable.
\\
% ab_arr = 20, column = 8, row = 9, bunch13, jumper was off - current draw 38 mA, Cf = 1966 fF
%Analysis code: C:\BIGSIS\KECK_APSPAD\Submission_2_DataAnalysis\Matlab_Image_Analysis\CHESS_image_analysis\Mon_Bunch14
%Data: C:\BIGSIS\KECK_TESTER2\KECK_APSPAD\Submission2_ReadoutControl\TestData\20100321_NBg3pg33_CHESS_E : bunch13
% att = 4
\indent
The results of Fig.~\ref{fig:everyTrainvsDelay} (and Fig.~\ref{fig:laserspeed}) indicate the feasibility of isolation of single bunches with around \unit[150]{ns} spacing, which is sufficiently fast for single-bunch imaging at the APS and ESRF (\unit[153]{ns} and \unit[176]{ns} separation, respectively).  The flat-top time in Fig.~\ref{fig:everyTrainvsDelay} of \unit[75]{ns} suggests that at CHESS the entire signal from the train would still have been
measured if the exposure time was reduced from \unit[190]{ns} to
\unit[115]{ns}. Further, at CHESS, the rising and falling edges of
Fig.~\ref{fig:everyTrainvsDelay} are slowed by the duration of the
trains; whereas, a single bunch is effectively a delta-function impulse at the detector.  The
reduction of the signal duration by \unit[56]{ns} should allow
reduction of the exposure time by at least \unit[25]{ns}.  Therefore, it is estimated that bunches at the APS could be resolved with an exposure time of \unit[90]{ns} and a time between frames for pixel reset of \unit[60]{ns}.\\
\indent
The results shown in Fig.~\ref{fig:everyTrainvsDelay} were used to evaluate the shift of the detector pedestal level and
read-noise induced by high levels of x-ray flux.  The time-averaged
flux at the brightest pixels was measured to be \unit[4.8$\times
10^{3}$]{x-rays/pix/$\mu$s} equivalent to \unit[2.1$\times 10^{11}$]{x-rays/mm$^2$/s}.  At PAD delays of \unit[2550-2600]{ns} storage elements $C_{S5}$ through $C_{S8}$ should have measured zero.  The read-noise of these, nominally empty storage elements, showed an insignificant maximum increase of $1.2\%$ at the highly illuminated pixels compared to images with the mechanical x-ray shutter closed.  The maximum and average integrated signal measured by the highly illuminated pixels in storage elements $C_{S5}$ through $C_{S8}$ were less than $<0.02\%$ and $<0.004\%$, respectively, of the signal measured by storage elements $C_{S1}$ through $C_{S4}$ (both of which could be due to random variation).\\
\indent
Due to an instability of the x-ray source (see below) it was not possible to use x-ray measurements to determine the time required to settle to high levels of accuracy.  In a separate set of experiments an in-pixel circuit injected \unit[235]{fC}, equivalent to 665 x-rays of \unit[8]{keV} energy, into the pixel input and the time to settle to within particular levels of the final value was measured.  The small-signal time constant and time for slew is longest for the smallest feedback capacitance; these times are reported here as a worst case condition.  The front-end amplifier transistor sizes were different on each half of the readout ASIC.  This resulted in slightly different speeds for the two sides so that two settling times are reported for every measurement.  At \unit[15]{$\mu$A} dissipation by the front-end amplifier, amenable to a larger array, 10-bit settling (within \unit[772]{$\mu$V} of the final value) required 100 and \unit[80]{ns} and 12-bit settling (within \unit[193]{$\mu$V} of the final value) required 120 and \unit[90]{ns}.  For the 16$\times$16 device per-pixel power dissipation is of less concern.  At a larger power dissipation, 12-bit settling (within \unit[193]{$\mu$V} of the final value) required \unit[65] and \unit[65]{ns}.  With the 16$\times$16 pixel device 12-bit settling was shown possible at times required to segregate bunches at the APS or ESRF.
\subsection{Beam characterization}\label{Sec:beam_charact}
\begin{figure}[] \centering
\includegraphics[width=3.5in]{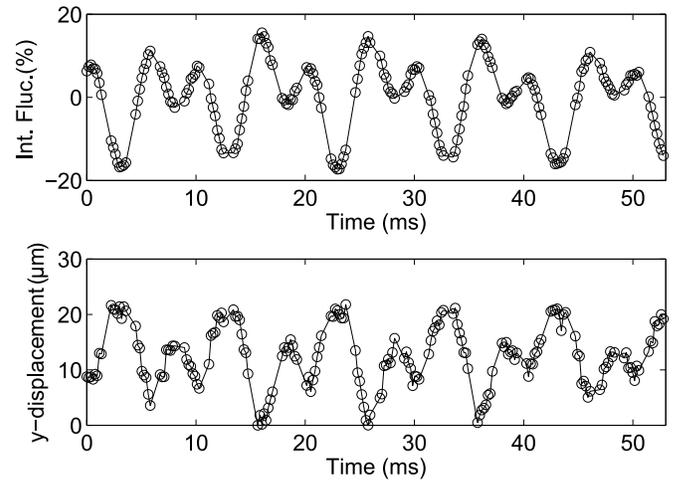}
\caption{Millisecond timescale studies of beam intensity fluctuations (top) and vertical displacement (bottom).  For this measurement the detector waited for the fifth train to circulate the synchrotron 75 times between captures of the in-pixel storage elements.  Lines guide the eye.}
\label{fig:IntFluc}
\end{figure}

\begin{figure}[] \centering
\includegraphics[width=3.5in]{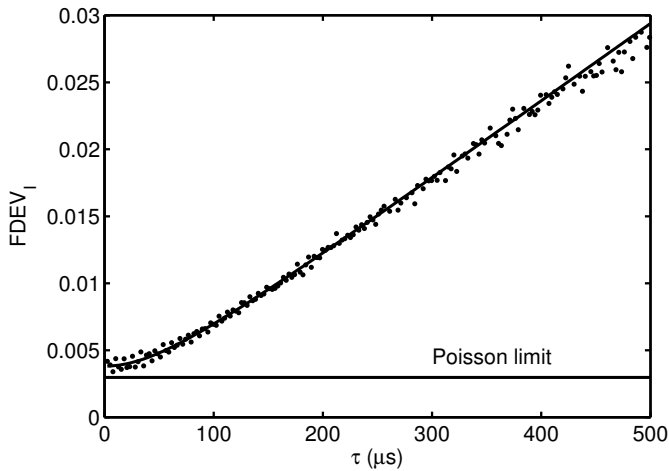}
\caption{Fractional RMS fluctuation of the CHESS G3
beam intensity versus correlation time.  The solid-line is a fit to
$FDEV_{I}^2 = A\tau^2+P$.}
\label{fig:RMSint}
\end{figure}

\begin{figure}[!htb] \centering
\includegraphics[width=3.5in]{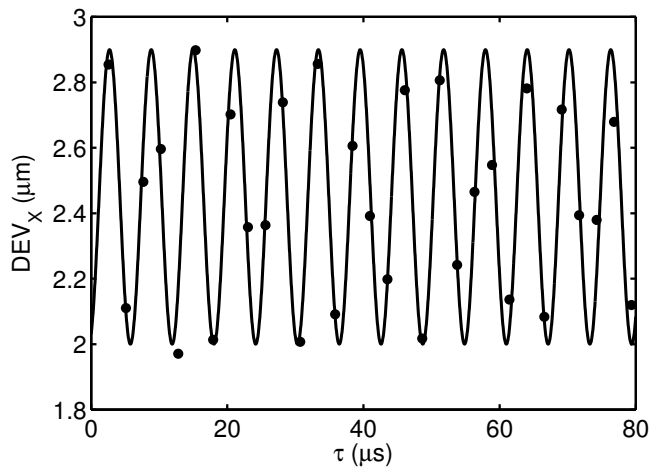}
\caption{Fluctuations of the CHESS G3 beam horizontal position at fast time-scales with a
superimposed oscillation (solid-line) at \unit[163]{kHz}.}
\label{fig:Xdev}
\end{figure}

The x-ray beam in the experimental hutch may have intensity and
position fluctuations due to trajectory instabilities of the
electron/positron source or from motions of the optics that interact
with the x-ray beam. Evaluation of these variations is important,
particularly for time-resolved experiments.  Diagnostics at or exceeding the
frequencies of vibrations that may be driven by pumps are valuable
for commissioning of x-ray beam lines.  \\
\indent
The fast readout time of the PAD camera and the
ability to resolve individual CHESS bunch trains allowed for
unique characterization of the x-ray beam on microsecond
time-scales.  Since the PAD directly images the beam on an array of
pixels the beam profile is extracted, which provides more diagnostic
information than four-quadrant type beam position
monitors.  The in-pixel storage was used to study temporal
correlations of a train sampled at each pass around the synchrotron
ring.\\
\indent To characterize the stability of the x-ray beam each in-pixel storage element of the PAD captured a unique single pass of train five.  The temporal separation between the capture of train five by each storage element was varied to study the time correlation of the position and intensity fluctuations.  Each image read out from the detector measured the x-ray signal at multiple time differences,
$\tau$, because of the eight in-pixel storage elements.  The time between PAD readout was limited to a minimum of \unit[700]{$\mu$s} and varied from \unit[719]{$\mu$s} to \unit[2.7]{ms} to extract a range of time correlations.\\
\indent
The intensity, $I$, was measured as the sum of the entire
spot; horizontal and vertical positions were found with a
center-of-mass algorithm.  The direct time evolution of the intensity and vertical position is shown in Fig.~\ref{fig:IntFluc}.  Oscillations were detected at \unit[100]{Hz} and \unit[200]{Hz} and were discovered to be induced by vibrations of the monochromator from a vacuum pump.
\\
\indent
To study faster time-scales, intensity and position time correlations were evaluated by calculation of the average RMS deviation versus the time difference, $\tau$, between the measurements.  This allowed for study down to the time of \unit[2.56]{$\mu$s} for the train to circulate the ring.  Fractional intensity deviations were evaluated as
\begin{equation} FDEV_{I}(\tau) =
\frac{\sqrt{<(I(t)-I(t+\tau))^2>}}{\sqrt{2}<I>},\end{equation} where $< >$ indicates an average.  Horizontal deviations were calculated similarly, but not normalized to the average position, and indicated as $DEV_{X}(\tau)$.  The integrated intensity measured per capture was around $1.1\times10^5$ x-rays which implies an accuracy limit from Poisson statistics for the intensity measure of $1/\sqrt{1.1\times10^5} = 0.0030.$\\
\indent
Fig.~\ref{fig:RMSint} displays the fractional intensity deviations extracted for correlation times from \unit[2.56]{$\mu$s} up to \unit[500]{$\mu$s}.  At the shortest correlation times the measurement is close to the accuracy limit set by Poisson statistics.   To show this, the measured values were fit to $FDEV_{I}^2 = A\tau^2+P$.  The parameters found were: $A = (5.8\times10^{-5}/[\mu s])^2$ and $P = (0.0038)^2$ where $P$ represents the deviation that is constant with correlation time and is bounded by Poisson statistics.  Extraction of a nearly Poisson limited measurement displays that the accuracy of the detector was maintained at high-flux levels and short exposure times.  The linear growth of the intensity deviation with correlation time proves that the intensity fluctuations of the single train were dominated by the \unit[100]{Hz} and \unit[200]{Hz} oscillations shown in Fig.~\ref{fig:IntFluc}.
\\
\indent
Fig.~\ref{fig:Xdev} displays the deviations of the center-of-mass of the x-ray intensity in the horizontal direction for correlation times up to \unit[80]{$\mu$s}.  The deviations oscillate at a frequency of \unit[163]{kHz} which is attributed to betatron oscillations of the circulating positron cloud.  The maximum sampling rate of this measurement was limited to the time for the train to circulate the ring.  As such, it is not possible to determine if the frequency extracted is the true frequency or an aliased measure of a higher frequency of the positron cloud motion.  The actual possible transverse oscillation frequencies, $f$, are then $ \unit[163]{kHz} = |f - N \times \unit[390.1]{kHz}|$ for any integer $N$  (the revolution frequency at CHESS is \unit[390.1]{kHz}).  Before these experiments the horizontal betatron measurements were measured by the storage ring operators at a frequency of \unit[227.7]{kHz}, which would be detected by the PAD as a frequency of $|\unit[227.7]{kHz} - \unit[390.1]{kHz}| = \unit[162.4]{kHz}$, consistent with measured results.
\\
\indent
These results show this detector to be appropriate for diagnostics of beam position and intensity fluctuations at, and exceeding, the typical frequencies of mechanical vibrations and many synchrotron ring instabilities.
\section{Discussion}
Pump-probe experiments use a laser to initiate dynamics in a sample which is then probed by an x-ray pulse with a known time delay with respect to the laser.  The detector described would be a useful tool for these experiments.  The detector readout time of \unit[600]{$\mu$s} is faster than the typical repetition rate of the pump laser.  The single-shot saturation value of the detector in a $\unit[1.2\times 1.4]{mm^2}$ spot was shown to exceed 100,000 x-rays.  This large capacity maximizes the accuracy of time-resolved measurements that are limited by photon statistics.  Another advantage is that, unlike a point detector, the two dimensional pixel array provides sensitivity to peak position and profile for detection of lattice parameter shifts and lattice inhomogeneity.  In-pixel storage elements may capture both the laser-on image and laser-off reference image in a single readout to reduce drifts.\\
\indent
This work will be leveraged for the planned development of a buttable, larger format device for a wider range of time-resolved experiments.  The pixel was designed with power dissipation appropriate for larger arrays.  The FPGA control code developed benefits future larger arrays since the majority of the state machines are independent of detector format.
\section{Conclusions}
A strength of analog integrating PADs is the high count-rate capability that minimizes uncertainty due to Poisson statistics at short exposure times.  In this paper an analog integrating area detector with speed sufficient for the segregation of the x-ray signal from a number of successive synchrotron pulses onto in-pixel storage elements was demonstrated.  The PAD was applied to studies of turn-by-turn x-ray bunch-train intensity and horizontal position measurements at correlation times down to \unit[2.56]{$\mu$s}.  Single Bragg spot time-resolved studies, especially of non-repetitive sample dynamics, are an excellent experimental match for future applications.\\
\section{Acknowledgments}
The authors acknowledge the Cornell detector group for their support: Darol Chamberlain, Katherine S. Green, Marianne S. Hromalik, Hugh Philipp, and Mark W. Tate; Martin Novak for machining of the enclosure; Arthur Woll, Tushar Desai, John Ferguson, and Katherine S. Green for help at the beamline.  We are also grateful to Tom Hontz and Area Detector Systems Corporation (ADSC) for the high-resistivity detector chips.  This work was supported by the William M. Keck foundation and DOE-BER Grant DE-FG02-97ER62443.  CHESS is supported by the US NSF and NIH-NIGMS through NSF grant DMR-0225180.

\end{document}